# Approaching the Purcell factor limit with whispering-gallery hyperbolic phonon polaritons in hBN nanotubes

Xiangdong Guo[1,3,#], Ning Li[2,#], Xiaoxia Yang[1,3,*], Ruishi Qi[2], Chenchen Wu[1,3], Ruochen Shi[2], Yuehui Li[2], Yang Huang[6], F. Javier García de Abajo[7,8,*], En-Ge Wang[4,5,9], Peng Gao[2,4,*], and Qing Dai[1,3,*]

[1]CAS Key Laboratory of Nanophotonic Materials and Devices, CAS Key Laboratory of Standardization and Measurement for Nanotechnology, CAS Center for Excellence in Nanoscience, National Center for Nanoscience and Technology, Beijing 100190, China.

[2]International Center for Quantum Materials, Electron Microscopy Laboratory, School of Physics, Academy for Advanced Interdisciplinary Studies, Interdisciplinary Institute of Light-Element Quantum Materials and Research Center for Light-Element Advanced Materials, Peking University, Beijing 100871, China.

[3]Center of Materials Science and Optoelectronics Engineering, University of Chinese Academy of Sciences, Beijing 100049, China.

[4]Collaborative Innovation Center of Quantum Matter, Beijing 100871, China.

[5]Songshan Lake Materials Lab, Institute of Physics, Chinese Academy of Sciences, Guangdong, China.

[6]School of Materials Science and Engineering, Hebei Key Laboratory of Boron Nitride Micro and Nano Materials, Hebei University of Technology, Tianjin 300130, China.

[7]ICFO-Institut de Ciencies Fotoniques, The Barcelona Institute of Science and Technology, 08860 Castelldefels (Barcelona), Spain.

[8] ICREA-Institució Catalana de Recerca i Estudis Avançats, Passeig Lluís Companys 23, 08010 Barcelona, Spain.

[9]School of Physics, Liaoning University, Shenyang, China.

[#]These authors contributed equally: Xiangdong Guo, Ning Li.

*emails: daiq@nanoctr.cn, p-gao@pku.edu.cn, javier.garciadeabajo@icfo.eu, yangxx@nanoctr.cn.

**Abstract:** Enhanced light-matter interaction at the nanoscale is pivotal in the foundation of nonlinear optics, quantum optics, and nanophotonics, which are essential for a vast range of applications including single-photon sources, nanolasers, and nanosensors. In this context, the combination of strongly confined polaritons and low-loss nanocavities provides a promising way to enhance light-matter interaction, thus giving rise to a high density of optical states, as quantified by the so-called Purcell factor –the ratio of the decay rate of an optical quantum emitter to its value in free space. Here, we exploit whispering-gallery hyperbolic-phonon-polariton (WG-HPhP) modes in hBN nanotubes (BNNTs) to demonstrate record-high Purcell factors ($\sim 10^{12}$) driven by the deep-subwavelength confinement of phonon polaritons and the low intrinsic losses in these atomically smooth nanocavities. Furthermore, the measured Purcell factor increases with decreasing BNNT radius down to 5 nm, a result that extrapolates to $\sim 10^{14}$ in a single-walled BNNT. Our study supports WG-HPhP modes in one-dimensional nanotubes as a powerful platform for investigating ultrastrong light-matter interactions, which open exciting perspectives for applications in single-molecular sensors and nanolasers.

**Introduction**

Light-matter interaction is the fabric of optics, permeating the manifestation of many phenomena such as reflection and refraction at macroscopic scales as well as stimulated emission from atoms and molecules at the microscale. As a result, the engineered enhancement of the light-matter interaction has led to several breakthrough applications including the development of nanolasers(*1, 2*), nano-biosensing elements(*3-5*), enhanced optical nonlinearities(*6, 7*), and cavity quantum electrodynamics(*8-10*). Consequently, the exploration of novel light-matter interaction regimes becomes a vital and highly desirable goal(*11, 12*). To this end, cavity-confined optical modes have proved to be instrumental, exhibiting a degree of interaction with light that is reflected in the density of optical states, which determines the increase in the decay rate of optical emitters relative to the rate in free space –the so-called Purcell factor(*13*).

For a confined optical mode, the Purcell factor is defined as the ratio of the quality factor ($Q$) to the optical volume ($V$)(*13*). Therefore, finding optical modes with an elevated Q factor and tighter confinement provides a route to increasing the Purcell factor. Plasmons constitute a viable tool to spatially confine the light field to the nanometer scale, thus permitting optical energy trapping in structures made of conducting materials. In addition, the mode lifetime can be increased by coupling these plasmonic structures to dielectric whispering-gallery microcavities, resulting in a significant enhancement in the quality factor due to the reduction of both radiation and scattering losses(*14-16*). In this context, highly doped two-dimensional graphene has been shown to exhibit relatively small plasmonic damping and strong optical spatial confinement(*17*), further increased by integrating this material in a vertical nanocavity to reach confinement down to one-atom spacing in the out-of-plane direction(*18-20*).

Compared to plasmons, hyperbolic phonon polaritons (HPhP) exhibit much lower losses and stronger light confinement, only limited by the atomic-vibration nature of these modes(*21-29*). For instance, the in-plane wavelength of HPhPs in monolayer hexagonal boron nitride (hBN) has been shown to reach values ~487 times smaller than the light wavelength $\lambda_0$ at the same infrared frequency(*30*). Furthermore, the HPhPs in nanostructured ~100 nm thick hBN flakes can be configured as cavity modes (e.g., in Fabry–Perot and photonic crystal geometries(*31, 32*)) to achieve a high three-dimensional mode volume compression ($V \sim 10^{-5} \lambda_0^3$). Inspired by these advances, we anticipate whispering-gallery HPhPs (WG-HPhPs)(*33, 34*) in small-volume nanocavities to push the Purcell factor to its ultimate limit. However, such extremely confined WG-HPhP modes have not been yet demonstrated due to the lack of supporting platforms.

Here, we experimentally demonstrate ultra-confined WG-HPhP modes propagating along hBN nanotubes (BNNTs), for which we measure a record-high Purcell factor (~$10^{12}$). By using electron energy-loss spectroscopy (EELS) in a scanning transmission electron microscope (STEM) to measure the WG-HPhP modes supported by BNNTs with different radii (~5-36 nm), we find an increase in the WG-HPhP Purcell factor with decreasing BNNT radius. In addition, a theoretical extrapolation of these results leads to a Purcell factor ~$10^{14}$ in a 1 nm radius single-walled BNNT (SWBNNT) –two orders higher than the one for 5 nm radius. We conclude that these WG-HPhP modes supported by one-dimensional nanotubes provide a powerful platform to realize strong light-mater interactions, and further offer a novel path for the development of ultra-miniaturized nanophotonic devices.

**Detection of WG-HPhP modes in BNNTs**

To detect the HPhPs modes in BNNTs, we employed monochromatic EELS incorporated in a STEM with Ångstrom spatial resolution and ~7.5 meV energy resolution operated at 30 kV and 60 kV (Fig. 1A, details in Methods). The BNNTs were grown by chemical vapor deposition methods (details in Methods), and their radii were mainly ~2-10 nm and ~20-50 nm (Fig. S1). To perform the STEM-EELS measurements, they were suspended on lacy carbon TEM grids, whose morphology with high-quality layered and hollow features is illustrated in the inset of Fig. 1A. The HPhPs were excited by the fast electrons travelling near the samples (aloof configuration)(*30*), resulting in distinct peaks in the STEM-EELS spectra, as shown in Fig. 1B. The details of the EELS data processing are described in Supplementary Note 1 and Fig. S2. Under aloof excitation, we identified the spectral features corresponding to the HPhP modes. HPhPs supported by hBN nanoflakes are propagating surface phonon polaritons(*30, 35*). Such modes are laterally confined in an hBN nanoribbon, giving rise to a single peak at ~173 meV in the EELS spectrum shown in Fig. 1B (black curve). In contrast, there are two peaks (~174 meV and ~192 meV) in the EELS spectrum of a BNNT (Fig. 1B, red curve). The additional peak in the BNNT configuration can be assigned to a whispering-gallery mode, which is scrutinized in what follows.

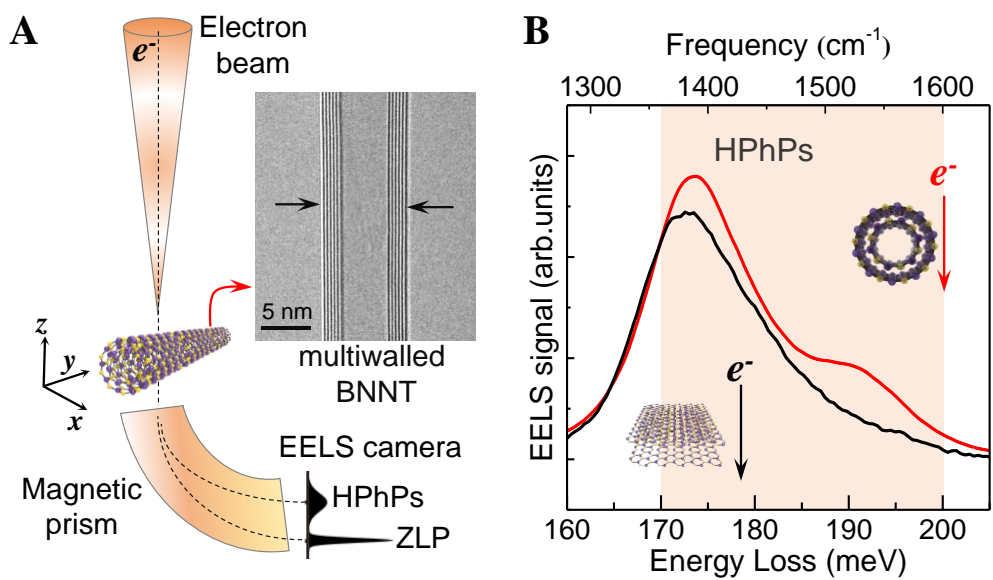

**Fig. 1. Fast electron excitation of HPhPs in BNNT.** (**A**) Schematic diagram of the electron energy-loss spectroscopy incorporated in a scanning transmission electron microscope (STEM-EELS). A focused electron beam (indicated by the orange cone) traverses a suspended BNNT and is subsequently collected by an EELS camera. The inset image illustrates the morphology of a multi-wall BNNT. (**B**) Measured spectra revealing multiple HPhPs in the BNNT under aloof excitation, in contrast to a single HPhP band in an hBN flake. The shaded area represents the upper Reststrahlen (RS) band of the hBN.

High-spatial resolution STEM-EELS provides an effective means to efficiently excite and detect highly confined polaritons. Moreover, the complex polariton signals can be identified and characterized by raster-scanning the electron beam over the spatial extension of the sample. Thus, we performed systematical STEM-EELS measurements by moving the electron beam along directions parallel (longitudinal) and perpendicular (transverse) with respect to the axis of the BNNT, as schematically illustrated in the inset of Figs. 2A and 2D, respectively. For a better depiction of the perpendicular scanning (with a ~3 nm spacing between the successive beam spots), a BNNT sample with an outer radius (R) of ~36 nm and an inner radius (r) of ~13 nm was studied. Under aloof excitation (i.e., with the electron beam passing through vacuum near the sample), the EELS spectra are well suited to probe pure HPhPs signals, while for bulk excitation (i.e., with the electron beam traversing the BNNT) both the HPhPs and the LO phonon signals are detected. There are three peaks (labeled as A, B, and C from lower to higher frequency) in Fig. 2A (longitudinal scan), while only two peaks (designated as D and E) show up in Fig. 2D (transverse scan).

To understand the observed spectral features, we perform finite element method (FEM) simulations of the STEM-EELS spectra (details in Methods). More specifically, a BNNT model with outer diameter $R$=36 nm, inner diameter $r$=13 nm and axial length $L$=1 μm (parameters taken from the measured BNNT image) is used. The employed dielectric tensor of the BNNT is described in cylindrical coordinates with radial and tangential components set to the out- and in-plane components of the dielectric tensor of an hBN film (Fig. S3). Besides the obtained agreement with EELS measurements (see below), the validity of this model is further supported by the experimentally measured Raman phonon signal, as well as the nano-infrared phonon signal of the BNNT (detailed information in Figs. S3-S5 and Supplementary Note 2). Fig. 2C,F divulge the resulting EELS calculations, which reveal the manifestation of a series of new resonance peaks, much narrower than those experimentally observed in Figs. 2A,D. This can be attributed to the finite energy resolution of the STEM-EELS. It should be noted that our experimental data has undergone deconvolution mathematical processing (details in Methods and Supplementary Note 2), so that the energy resolution is approximately from 7.5 meV to 5 meV. Therefore, we convolute the calculated EELS signals with a Gaussian function of 5 meV full width at half maximum (FWHM) to obtain the spectra presented in Figs. 2B,E, which are in excellent agreement with the respective experimental spectra (Figs. 2A,D).

In the EELS longitudinal scan with aloof excitation (Fig. 2A-C), we interpret the observed resonant features (A, B and C peaks) as HPhP modes, similar to those in nanoflakes(*30, 35*), but strongly modified by the small size and cylindrical warping of the BNNT. As the electron beam is moved from the end to the middle of the nanotube along the axial direction, the frequencies of both peaks A (~172 meV) and C (~193 meV) remain almost unchanged, while that of peak B (172 meV-193 meV) monotonically decreases. Consequently, peak B is a characteristic longitudinal Fabry-Perot (FP) resonance in the BNNT, as can be further confirmed upon inspection of the EELS calculations before convolution (Fig. 3C). Indeed, in the 172-193 meV energy range (green shaded area), there are many resonant peaks whose spectral positions do not vary along the scan, and only the intensities of such peaks are changing. We must underline that the peak intensities strongly depend on the electron beam position because the excited HPhPs propagate along the BNNT toward the ends, where they are reflected, thus producing a characteristic interference pattern. This behavior clearly indicates a spectrally evolving standing wave, which is the longitudinal HPhP modes. Interestingly, this effect is comparable to the HPhP resonances in the hBN rods(*31*). Thus, the multi-peaks that are revealed in Fig. 2C can be regarded as many multi-level longitudinal HPhP FP-cavity resonant modes. In addition, the signal for peak A (Fig. 2A) mainly originates from the lowest-order mode in the multi-level resonance signal (Fig. 2C), which can also be detected far away from the end of BNNT in a vacuum (Fig. S6). Furthermore, peak B merges with peak A as the electron beam moves to the middle of the nanotube, implying that the physical origins of peak A and peak B are the same and both belong to the fundamental surface HPhP mode (SM0) in the BNNT. On the other hand, the signal of peak C (Fig. 2A) stems from the feature with the highest frequency before convolution (Fig. 2C), whereas its frequency is close to the surface optical (SO) phonon (~195 meV). We can also observe that peak C basically exceeds the limit frequency of the longitudinal HPhP modes. Thus, it is anticipated that peak C is the transverse WG-HPhP of BNNT.

The transverse scan (Fig. 2D) provides further insight into the WG-HPhP modes. In the aloof excitation region, the two observed peaks (D and E in Fig. 2D) can be ascribed as FP and WG-HPhP modes. In particular, the FP feature converges to the SM0 peak when the electron beam reaches the center of the BNNT. Although the energies of these two peaks hardly change along the scan, their intensities increase as the electron beam approaches the BNNT. However, during the bulk excitation, the FP mode remains unchanged while the high-frequency peak shifts to blue as the electron beam moves into the inner region of the BNNT. As expected, the transverse scan renders symmetric results with respect to the nanotube axis. Such behavior indicates a spectrally evolving standing wave pattern along the direction perpendicular to the axis coming from the transverse WG-HPhP mode of the BNNT. In addition, we note that the high-frequency peak contains contributions from both the WG-HPhP modes and the longitudinal optical (LO) phonon (~200 meV) signal, as far as the bulk excitation is concerned.

The simulated EELS spectra before convolution allow us to resolve the WG-HPhP modes and the LO phonon in Fig. 2F. There are three peaks in the high-energy region (pink shaded area), which are identified as many multi-level transverse WG-HPhP resonance modes. Interestingly, one of the peaks at ~198 meV (around the LO phonon) changes significantly in spectral position and shape, implying that a superposition effect of at least two peaks (WG-HPhP mode and LO phonon) should take place. Additionally, its frequency gradually red-shifts, while the signal intensity gradually decreases until it becomes zero as the electron beam moves from the center to outside of the BNNT. The other two resonant peaks (~192 meV and 196 meV) are the pure WG-HPhP modes, whose frequencies do not vary and only their intensities change across the transverse scan. From the above analysis, the peak E of Fig. 2D can be confirmed as a WG-HPhP. The whispering gallery mode on an optical nanocavity system has been directly observed in our experiments, which has been pursued in the process of light-matter interaction for the past decade.

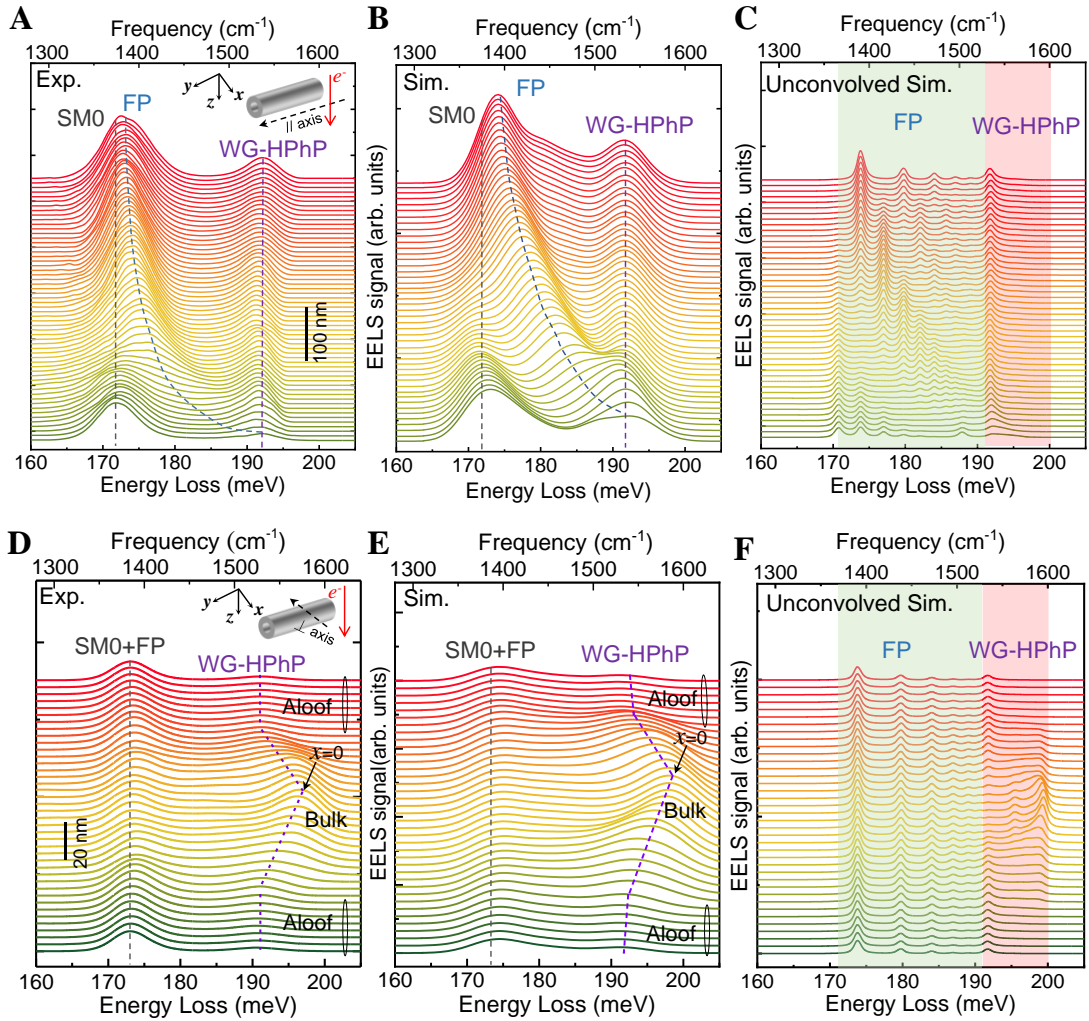

**Fig. 2. EELS characterization of hyperbolic polaritons in a BNNT. (A, D)** Series of EELS spectra obtained when the electron beam is scanned along the axis of the BNNT (in the $y\sim 0$ to 500 nm range, with $x=42$ nm, see inset) (A) and perpendicular to the axis of the BNNT (in $x\sim -54$ to 54 nm, with $y=500$ nm) (D). The BNNT has a radius of 36 nm and is parallel to the $y$ direction. **(B, E)** Simulated EELS spectra after convolution with a Gaussian function considering the broadening effect in the experiment data in (A) and (D). **(C, F)** Unconvolved simulated EELS spectra corresponding to (B) and (E). The green and pink shaded areas indicate the Fabry-Perot modes and the WG-HPhP modes. Insets to (A), (D) show the electron beam (downward red arrows) and scanning (dashed arrows) directions.

**Dispersion and modulation of WG-HPhPs in BNNTs**

To quantify the modulation of the HPhPs mode signals by electron beams in different spatial positions, we investigated both the unconvolved EELS simulations and the corresponding near-field electromagnetic mode distributions in the BNNT. Figure 3A shows two typical unconvolved EELS simulations when the electron beam is raster-scanned along the *x*-axis (perpendicular to the nanotube axis) from $x=0$ (BNNT core, blue curve) to $x=R$ (BNNT surface, red curve), with the longitudinal resonances denoted as peak I-V and the two WG-HPhPs resonances denoted as peak VI and VII. Their respective electromagnetic field distributions are exhibited in Fig. 3B,C. The most apparent difference between the WG-HPhP and FP-cavity modes is that the electromagnetic field oscillates in the plane perpendicular to the nanotube for the WG-HPhPs (Fig. 3B) and along the axis direction for the FP-cavity modes (Fig. 3C). In addition, the local field distribution of the FP modes on the annular section of the BNNT is uniformly confined on the surface of the BNNT (no interference effect, Fig. 3B), contrasted to the WG-HPhP modes with apparent cyclic transmission (Fig. 3C). This further corroborates that the WG-HPhPs propagate circularly around the nanotube during the manifestation of the FP modes along the nanotube. The propagating distance between the two nearest near-field signal maxima is half the wavelength of the polaritons. For either the FP- or WG- HPhPs modes, with the increase of the order in HPhPs (e.g. from the peak I to V for the FP modes, or from peak VI to VII for the WG-HPhPs modes), the number of the near-field signal maxima can increase, and the wavelength of the polariton becomes shorter and the wavelength compression is stronger. However, as can be ascertained, the wavelength compression of the WG-HPhPs (peaks VI and VII) can exceed the limits of FP modes (peaks I-V) due to its extremely small volume.

The excited WG-HPhP modes can be selected through the position of the incident electron beam. More specifically, at the BNNT core ($x=0$), the excited WG-HPhP mode is symmetric to the external electron beam (peak VII in the blue curve of Fig. 3A), in sharp contrast to the electromagnetic distribution for $x=R$ (peak VI in the red curve of Fig. 3A). The main difference is that when the electron beam moves into the BNNT, it interacts twice with the nanotube wall, thus introducing two excitation sources at the top and bottom sides. When *x*=0, four near-field signal maxima are generated on the ring cavity due to the existence of two excitation sources, while the corresponding interference wavelength of the WG-HPhPs is approximately half the circumference ($\lambda_1 = \lambda_{x=0} \simeq \pi(R+r)/2$, in the blue curve). On the other hand, when *x*=R, only one equivalent single direct electron excitation source can generate two near-field signal maxima on the ring cavity. The resulting interference wavelength of the WG-HPhPs is approximately the entire circumference ($\lambda_2 = \lambda_{x=R} \simeq \pi(R+r)$, in the red curve). Both peaks of VI and VII are the fundamental electromagnetic modes.

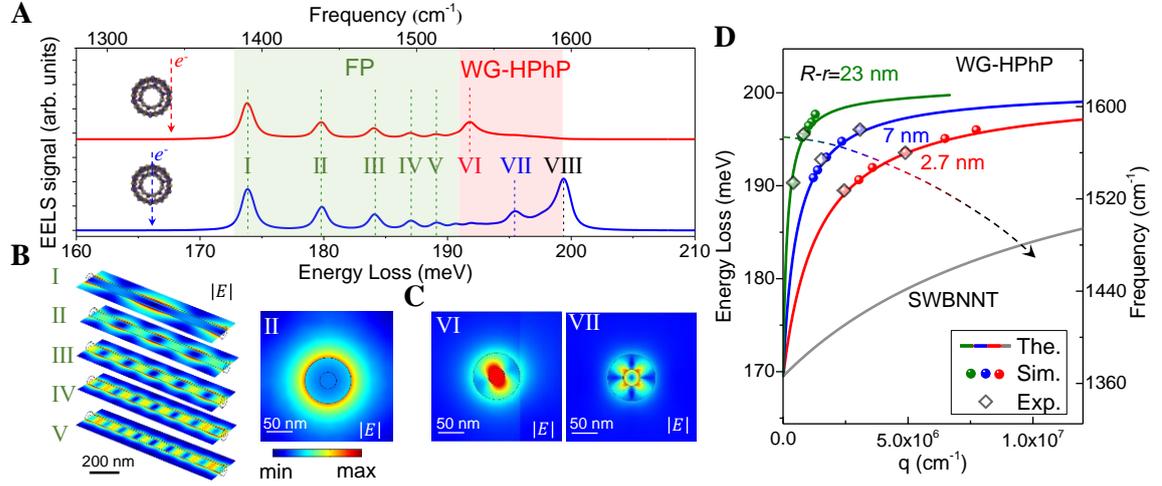

**Fig. 3. Polariton mode confinement and dispersion relationship in a BNNT. (A)** Simulated EELS spectra when the electron beam passes grazingly (red curve) or through the center (blue curve) of the BNNT (36 nm radius), respectively. **(B, C)** Near-field distribution of the optical mode at the resonant peaks in the EELS spectra. We present the value of |**E**| for peaks I-V (FP modes) in (B), and for peaks VI-VII (WG-HPhP modes) in (C). **(D)** Dispersion relation of the WG-HPhP modes in BNNTs of different sidewall thicknesses (external minus internal radii): 23 nm (green), 7 nm (blue), 2.7 nm (red) and 0.34 nm (SWNNT, black); solid curves stand for theoretical model calculations; balls are numerical simulations from Fig. 3A and Fig. S9; diamonds are measured EELS data from Fig. 2D and Fig. S8.

To quantify the modulation of polaritons, we analyzed the dispersion relation of both the WG-HPhP and FP modes. The pseudo-colour relation along the axis (dispersion from the experimental data, Fig. S6 and Supplementary Note 3) can directly illustrate the FP mode that agrees well with the calculated dispersion (Fig. S6 and Supplementary Note 3). As far as the WG-HPhP modes are concerned, their dispersion cannot be extracted experimentally since the momentum cannot be directly estimated by the position of the electron beam perpendicular to the nanotube (Fig. S7). Moreover, the electromagnetic energy of the WG-HPhP modes is mainly distributed inside the BNNT, which is similar to the volume-HPhP modes in hBN films and different from the longitudinal surface-HPhP FP mode of the BNNTs. Hence, WG-HPhP modes can be understood as HPhP modes propagating along an hBN film wrapped into a circular tube surface. This idea suggests the analytical expression $q(\omega) + ik(\omega) = -\frac{\psi}{t}[2\arctan\left(\frac{\varepsilon_0}{\psi}\right)]$, where $q$ and $k$ are the real and imaginary parts of the HPhP wave vector, $\varepsilon_0$ is the free-space permittivity, the ratio $\psi = \sqrt{\varepsilon_\parallel}/i\sqrt{\varepsilon_\perp}$ involves the dielectric function of the BNNT along tangential ($\varepsilon_\parallel$) and radial ($\varepsilon_\perp$) directions, and $t$ is sidewall thickness of the BNNT(*21*). To verify the above-mentioned analytical dispersion model for the WG-HPhP modes, we analyze multiple sets of EELS spectra for BNNTs with sidewall thicknesses (R-r) of 23 nm, 7 nm, 2.7 nm by performing both experiments and FEM-based simulations. The experimental resonant frequency-wave vector values of the WG-HPhP modes (balls in Fig. 3D) are extracted from the EELS data in Fig. 2D and the Fig. S8 (details in Supplementary Note 3). We also simulate EELS spectra for the series of BNNTs with different radii, as well as sidewall thicknesses (Fig. S9). According to $2\pi/q_{x=0} = \lambda_{x=0} \simeq \pi(R+r)/2$, we extract the frequency and the wave vector of the WG-HPhP mode excited at the position $x = 0$ from the simulated EELS spectra and plot them as balls in Fig. 3D. In addition, diamonds in Fig. 3D are measured EELS data from Fig. 2D and Fig. S8. These values are mutually consistent. On the basis of this analysis, we theoretically predict the dispersion relation for a SWBNNT of ~0.34 nm thickness in Fig. 3D.

From the obtained dispersions, we see that the resonance frequency and wave vector $q$ of the WG-HPhP modes can be effectively modulated by varying the size of the BNNT. For a specific frequency, the $q$ value will increase by decreasing the BNNT sidewall. Moreover, for a given BNNT sidewall thickness, the dispersion distribution of the WG-HPhP mode is determined (i.e., $q$ will increase as the BNNT radius declines). Therefore, smaller BNNTs can support the more confined WG-HPhP modes, implying that the highest degree of spatial confinement can be achieved in SWBNNTs.

**Purcell factor limit of WG-HPhP modes**

By considering that the BNNT's WG-HPhP modes exhibit high wavelength compression and enhanced optical field confinement capabilities, it is possible to reach unprecedented values of the density of optical states, as quantified by the Purcell factor $P \sim Q/V_m$, where $Q$ is the quality factor and $V_m$ is the effective mode volume $V_p$ normalized to the cube of the free-space light wavelength $\lambda_0^3$.

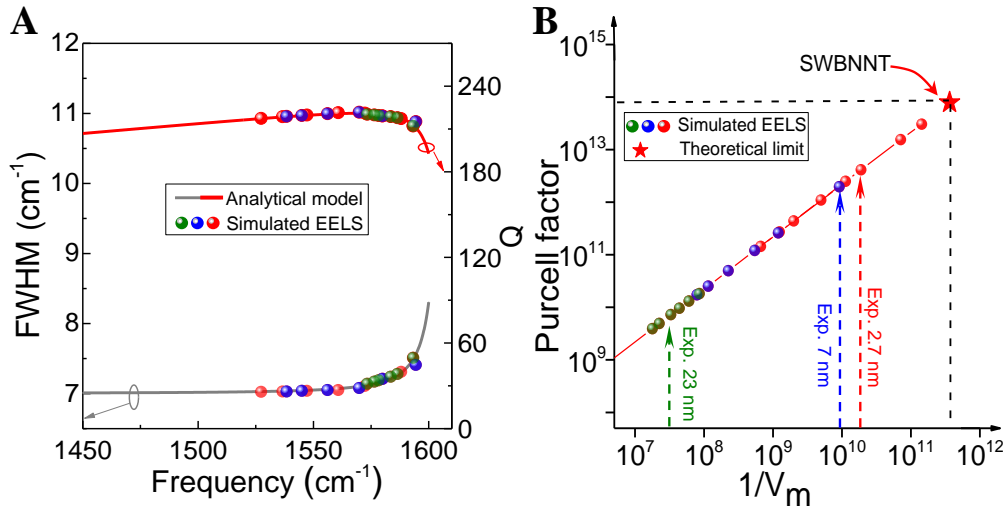

**Fig. 4. Record-high Purcell factor of WG-HPhP modes in a BNNT.** (**A**) Spectral width (FWHM, left vertical scale) and quality factor ($Q$, right scale) of the WG-HPhP modes obtained from the unconvolved EELS simulations corresponding to the experimental data of BNNT with different sidewall thicknesses (23 nm, 7 nm, and 2.7 nm) in Fig. 3D (balls). They agree well with the analytical model results (solid curves). (**B**) Ultra-small volume and Purcell factor of WG-HPhP modes in BNNTs. A record-high Purcell factor is predicted for a SWBNNT. The data points are calculated from Figs. 3D and 4A, so we retain the same color code. The experimental maximum values of 1/$V_m$ and Purcell factor in BNNT with sidewall thickness (23 nm in Fig. 2, 7 nm in Fig. S8, and 2.7 nm in Fig.3 and Fig. S8) are marked in (B).

The quality factor of the WG-HPhPs can be extracted from the EELS spectra as Q=$\omega_{res}$/γ, where $\omega_{res}$ is the resonance frequency and γ is the FWHM of the EELS peak. Therefore, we perform Lorentz fitting and extract the FWHM of the WG-HPhP resonant peaks in the EELS spectra that are revealed in Fig. 3A and Figs. S10 and S11 (details in Supplementary Note 4). The results are plotted in Fig. 4A (dots with color matching the dispersion in Fig. 3D). The FWHM gives the loss rates of the HPhP modes, which contain three main damping pathways: intrinsic propagation losses, radiation losses and inelastic edge scattering. For the WG-HPhP modes that are supported in BNNTs, a closed structure system without boundaries, the edge scattering is eliminated. Subsequently, we calculated the propagation loss $\gamma_p$ of the WG-HPhP modes by enforcing the following analytical model(*36*): $\gamma_p = 2|k|v_g$, where $v_g = \partial\omega/\partial q$ stands for the group velocity. The calculated propagation loss is plotted as a function of the frequency in Fig. 4B, while the outcomes are consistent with the acquired FWHM of the simulated intrinsic absorption spectra of the BNNT (Fig. S10). On top of that, the propagation losses are in good agreement with the FWHM of the simulated EELS results. This indicates that the radiation losses can be ignored for the WG-HPhP modes in BNNTs, as expected from the ultra-high spatial confinement (i.e., the dramatic mismatch between the nanotube width and the light wavelength). Therefore, the intrinsic propagation losses (e.g., scattering by material defects and isotopic inhomogeneities) provide the only damping pathway of the WG-HPhP modes. These low losses result in high-quality factors (~220), as depicted in Fig. 4A. However, we have to underline that they are recognized within a 5 nm radius BNNTs, which overcomes the tradeoff between the mode volume and the polariton losses.

We now calculate the effective optical mode volume of the WG-HPhPs inside BNNTs via $V_m = V_p/V_0 = V_p/\lambda_0^3$. By taking into account that the fields decay by a factor of $1/e$ at a distance of $\lambda_{\text{WG-HPhP}}/2\pi$ in the direction perpendicular to the surface of the BNNT, the mode volume of the WG-HPhP can be estimated as follows: $V_p \sim \lambda_{\text{WG-HPhP}} \times (\lambda_{\text{WG-HPhP}}/2\pi)^2$, so we have $V_m = \lambda_{WG-HPhP}^3/4\pi^2\lambda_0^3$, which can be derived from the dispersion relationship in Fig. 3B. Based on the obtained values of $Q$ and $V_m$ values, we calculated the Purcell factors ($P=Q/V_m$) and plot them as a function of $1/V_m$ in Fig. 4B. Since the mode confinement ($\lambda_0/\lambda_{\text{WG-HPhP}}$) is proportional to $1/V_m$, we can observe that $P$ increases continuously as the confinement increases in BNNTs. Additionally, due to the ultra-small mode volume ($1/V_m \sim 10^{10}$) of the WG-HPhP modes, which can also have a high-quality factor resulting from the low intrinsic loss, $P$ can reach values of $\sim 10^{12}$, which demonstrates a quite strong light-matter interaction ever found experimentally. Also, from the acquired data, our model predicts an extreme *P-value* of $\sim 10^{14}$ for a SWBNNT (Fig. S12).

These extraordinary characteristics are derived from both the intrinsic material properties, as well as the employed 1D nanocavity structure. Compared to the plasmons with ohmic losses, phonon polaritons possess much lower inherent losses. BNNT is a natural whispering gallery nanocavity, which results in negligible edge scattering and radiation losses. Moreover, there is no optical phase change in the propagating WG-HPhP modes in BNNTs. These characteristics are ideal to maintain an excellent quantum coherence and a long quantum coherence time, thus providing a promising platform for realizing a single-photon source for quantum computing and quantum information.

In conclusion, we demonstrate record Purcell factors ($\sim 10^{12}$) that are driven by the natural WG-HPhP modes of a BNNT with a radius of $\sim 5$ nm. This BNNT constitutes the smallest whispering-gallery mode nanocavity that has been so far reported ($\sim 10^{-10}$ reduction in optical volume), with an ultra-high quality factor ($\sim 220$). These WG-HPhP modes are efficiently excited and detected with high-resolution monochromatic EELS, and identified by finite element electromagnetic field simulations with optical dielectric functions supplied by nano-FTIR measurements. Our study provides a new paradigm, which is anticipated to trigger further research on the optical properties of one-dimensional nanomaterials. Besides, the observed Purcell factors reveal extraordinary strong light-matter interactions, holding high potential for long-sought-after applications in both single-quantum emission and single-molecule detection at the atomic scale.

## Methods:

**Sample preparation for the TEM experiments.** The thick BNNT samples (radii ~20-50 nm) and thin BNNT samples (radii ~2-10 nm) were synthesized using the previously reported method.(*37, 38*) The final BNNT products were then dispersed in pure Ethanol in an ultrasonic oscillator for 30 minutes separately. Consequently, three droplets of each solution were transferred onto the 3 mm lacy carbon TEM grid. Before the STEM experiments, the samples were annealed at 160 °C for 8 hours in a vacuum chamber, to remove possibly existing hydroxide contamination.

**EELS and imaging experiments.** The EELS experiments were carried out on a Nion U-HERMES200 electron microscope with a monochromator that operated at 60 kV to avoid any damage to BN materials. We employed a convergence semi-angle $\alpha = 20$ mrad and a collection semi-angle $\beta = 25$ mrad for all datasets. In this setting, the spatial resolution was ~0.2 nm, while the energy resolution is ~7.5 meV, suitable for the characterizations of BNNTs. Moreover, the beam current used for EELS was ~10 pA, while the acquisition times were 200 ms/pixel and 500 ms/pixel for the datasets along with parallel and perpendicular directions with respect to the axis of the BNNTs, separately. The zero-loss peak (ZLP) was slightly saturated to improve the signal-to-noise ratio of the spectra. The HRTEM images in Fig. 1A and Fig. S1D-F were obtained using an FEI Tecnai F20 operated at 200 kV. Additionally, the high angle annular dark field (HAADF) image in Fig. 1A was obtained using a FEI Titan Cube Themis operated at 300 kV with a convergence semi-angle of 30 mrad and a collection semi-angle of 80 to 379 mrad.

**EELS data processing.** All the acquired vibrational spectra were processed by using the custom-written MATLAB code and Gatan Microscopy Suite. More specifically, the EEL spectra were firstly aligned by their normalized cross-correlation and then normalized to the intensity of the ZLP. Subsequently, the block-matching and 3D filtering (BM3D) algorithms were applied for removing the Gaussian noise. The background arising from both the tail of the ZLP and the non-characteristic phonon losses was fitted by employing the modified Peason-VII function with two fitting windows and then subtracted in order to obtain the vibrational signal. The Lucy-Richardson algorithm was then employed to ameliorate the broadening effect induced by the finite energy resolution, taking the elastic ZLP as the reference point for the spread function. The spectra were summed along the direction parallel to the interface for obtaining line-scan data with a good signal-to-noise ratio. In addition, we employed a multi-Gaussian peak fitting method to extract the polariton peaks from the composed signal.

**Theoretical calculations.** We employ a finite element method implemented in Comsol Multiphysics to simulate the EELS spectra. The *Radio Frequency Toolbox* is used for performing retarded simulations (solving Maxwell's equations) to evaluate the electric field in the presence of a BNNT. The dielectric function of the BNNT needed to calculate the optical response is detailed in Fig. S3 and Supplementary Note 2.

By following a well-established procedure(*39*), a current source is used to represent the electron beam along the direction perpendicular to the sample(*40-42*),

$$\mathbf{j}(\mathbf{r},\omega) = -e\hat{\mathbf{z}}\delta[\mathbf{r} - \mathbf{r_0}]e^{i\omega z/v} \qquad (1)$$

where the electron (a point charge $-e$) that moves with constant velocity $v$, hitting the sample at the position $\mathbf{r_0} = (x_0, y_0, 0)$, and we work in angular-frequency space $\omega$. The fast electron imposes an external evanescent field $\mathbf{E_0}(\mathbf{r},t)$ as it moves in vacuum. When it strikes on the sample, an induced field $\mathbf{E}^{\text{ind}}(\mathbf{r},t) = \mathbf{E}(\mathbf{r},t) - \mathbf{E_0}(\mathbf{r},t)$ is generated, where $\mathbf{E}(\mathbf{r},t)$ is the total field distribution, as determined by the optical response of the BNNT. Thus, the calculation for each frequency value is performed twice: with and without the existence of the BNNT, preserving the same mesh. After the solution for the electric field is obtained, we calculate the loss probability (by using *Edge Probe, Integral* applied along the electron trajectory) according to(*35, 39*)

$$\Gamma_{\text{EELS}}(\omega) = \frac{e}{\pi\hbar\omega}\int \text{Re}\left\{\exp\left(-\frac{i\omega z}{v}\right) E_z^{\text{ind}}(z,\omega)\right\} dz \qquad (2)$$

Therefore, the EELS spectra are simulated according to Eq. (2) for electrons passing the BNNT and the extracted outcomes are presented in Figs. 2C,F and Fig. 3A. For a quantitative comparison between the experiment and the theory, the spectral resolution determined by the measured ZLP needs to be considered. Thus, we convolve the calculated spectra of Figs. 2C,F with a Gaussian distribution of full FWHM ~5 meV:

$$\Gamma_{EELS}^{\text{broadened}}(\omega) = \int d\omega' \, \Gamma_{EELS}(\omega - \omega') \, \text{ZLP}(\omega') \tag{3}$$

Finally, we compute the EELS probability for different locations of the electron beam, as it is displayed in Figs. 2B,E, obtaining excellent agreement with the respective experimental data.

## Acknowledgements

This work was supported by the National Natural Science Foundation of China (51925203, 52022025, 52102160, 51972074, 11674073, 11974023, 52021006, and U2032206), the Key Program of the Bureau of Frontier Sciences and Education, Chinese Academy of Sciences (QYZDB-SSW-SLH021), the Strategic Priority Research Program of the Chinese Academy of Sciences (XDB30000000 and XDB36000000), the Key Research Program of the Chinese Academy of Sciences (ZDBS-SSW-JSC002), Youth Innovation Promotion Association C.A.S., C.A.S. Interdisciplinary Innovation Team (JCTD-2018-03), the '2011 Program' from the Peking-Tsinghua-IOP Collaborative Innovation Center of Quantum Matter. We also acknowledge the Electron Microscopy Laboratory of Peking University for the use of electron microscopes. F.J.G.A. acknowledges support from ERC (Advanced Grant 789104-eNANO), Spanish MCINN (PID2020-112625GB-I00 and CEX2019-000910-S), and Catalan CERCA Program.

## Data availability

The data that support the findings of this study are available from the corresponding author upon reasonable request.

## Author contributions

The concept for the experiment was initially developed by Q.D., X.Y., and P.G. Hexagonal BN nanotubes were grown by H.Y. STEM–EELS and TEM imaging experiments were performed by N.L. under the direction of P.G. and E.W. PTIR experiments were performed by X.G. and C.W. FEM simulations and theoretical analysis were performed by X.G. under the supervision of Q.D., X.Y., and F.J.G.A.

Data processing and analysis were performed by N.L. and X.G. assisted by R.Q., Y.L., and R.S.; X.G., X.Y., and N.L. wrote the manuscript with input from P.G., F.J.G.A., and Q.D.; X.G. and N.L. contributed equally to this work. All authors discussed the results at all stages and participated in the development of the manuscript.

## Competing interests

The authors declare no competing financial interests.